\documentclass[11pt]{article}

\usepackage{acl}

\usepackage{times}
\usepackage{latexsym}

\usepackage[T1]{fontenc}

\usepackage[utf8]{inputenc}

\usepackage{microtype}

\usepackage{inconsolata}

\usepackage{graphicx}

\title{CHICO: Bridging Sparse Metadata and Multimodal Evidence in Cultural
Heritage Collections}

\author{
  {\normalfont\bfseries
    Vagner Inácio de Oliveira, Paula Dornhofer Paro Costa} \\
  {\normalfont\ttfamily
    v208939@dac.unicamp.br, paulad@unicamp.br} \\[0.8em]
  Universidade Estadual de Campinas (UNICAMP) \\
  Faculdade de Engenharia Elétrica e de Computação (FEEC) \\
  Campinas, SP, Brazil
}

\begin{document}
\maketitle
\begin{abstract}
Cultural heritage aggregators provide access to millions of digitized items. Yet their catalog records often contain sparse, heterogeneous metadata, with associated media links that may not lead to accessible images.
Structuring and validating these resources for multimodal AI research can support reproducible experimentation.
To this end, we present CHICO (Cultural Heritage Images and Catalog Objects), a multimodal dataset derived from Europeana that brings together curated catalog fields, images, and contextual information to support evidence-grounded description generation while retaining the original Europeana records.
It comprises more than 60{,}000 cultural heritage records with associated images.
Although originally developed for that purpose, it can also support multimodal information retrieval, metadata enrichment, and evidence-based evaluation.
\end{abstract}

\section{Introduction}
\label{sec:intro}

Museums, libraries, and archives have digitized millions of cultural heritage items, making their collections increasingly accessible through aggregators such as Europeana~\citep{isaac2013europeana} and Brasiliana Museus~\citep{siqueira2022brasiliana}.
These collections bring together records from different institutions, reflecting diverse cataloging practices, metadata structures, and completeness levels.
In this context, catalog descriptions help visitors discover and understand cultural heritage items, while their availability and level of detail reflect the diversity of collections and cataloging practices.
Alongside catalog metadata, associated images and contextual notes can offer complementary perspectives.
Together, these sources enable multimodal approaches, including generating textual descriptions grounded in the information associated with each item.

The availability and structure of these information sources vary considerably across cultural heritage records.
Catalog information, including titles, dates, and creators, is documented at different levels of detail across collections.
Similarly, media links do not always provide access to usable images.
Institutional descriptions, when available, range from brief statements to detailed curatorial texts and may contain information not represented in other metadata fields or images.
Consequently, treating these descriptions as ground truth for generation or evaluation may introduce information that cannot be verified against the available evidence.
Furthermore, Europeana records follow the Europeana Data Model (EDM), combining nested structures, multilingual values, and URI-based references~\citep{isaac2013europeana}.
Bringing these information sources together for multimodal approaches involves selecting relevant catalog fields, handling heterogeneous metadata, and verifying the usability of associated images.

CHICO (Cultural Heritage Images and Catalog Objects) is a multimodal dataset derived from Europeana records to support evidence-grounded description generation.
It brings together curated catalog fields, associated images, and contextual information while preserving the original Europeana records.
Its structure provides a consistent representation of these complementary sources, supporting reproducible experimentation and potential reuse in other multimodal approaches.
The dataset comprises 60{,}397 cultural heritage records from 74 providers, with 83{,}852 images verified as decodable.
Construction and analysis code are publicly available;\footnote{\url{https://github.com/AI-Unicamp/heritage-data-hub}} the dataset artifacts are released on Zenodo~\cite{oliveira2026chico}.\footnote{\url{https://doi.org/10.5281/zenodo.23954726}}

\begin{figure*}[t]
    \centering
    \includegraphics[width=0.90\textwidth]{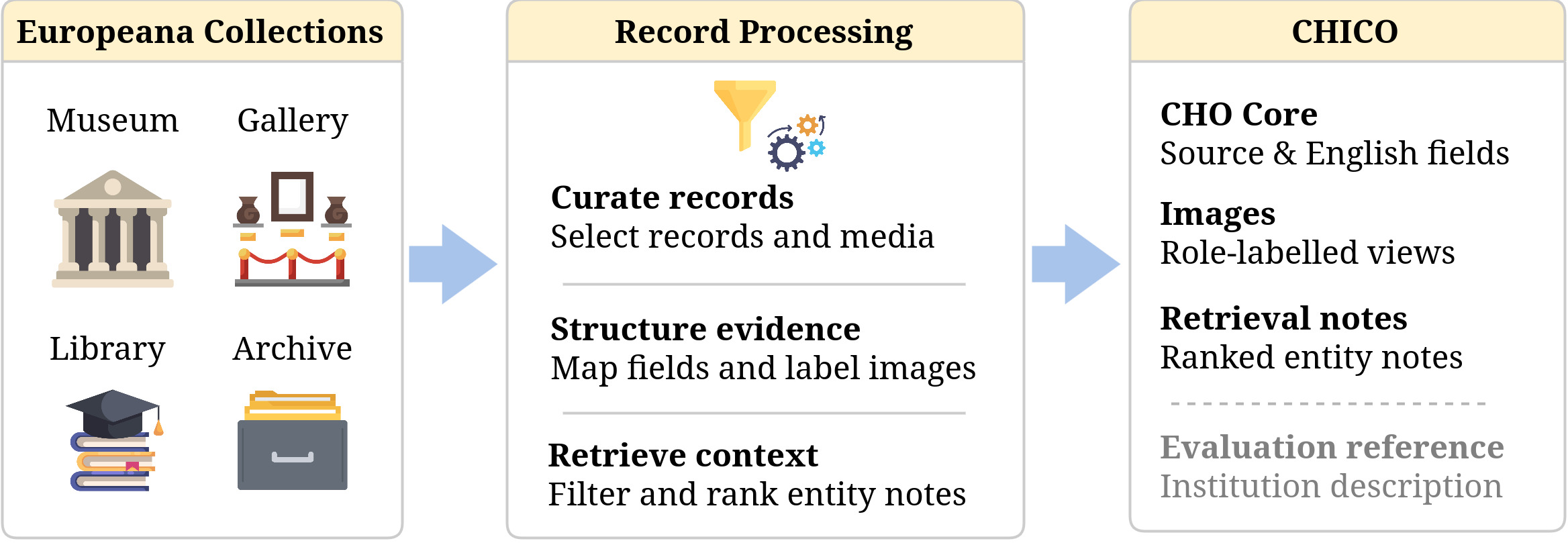}
    \caption{Construction of CHICO from Europeana collections into catalog fields, role-labeled images, and ranked retrieval notes, with the institution description retained as an evaluation reference.}
    \label{fig:chico-pipeline}
\end{figure*}

Figure~\ref{fig:chico-pipeline} illustrates the construction of CHICO, from Europeana collections to the resulting dataset.
The process involves curating records and associated media, structuring catalog information alongside images, and incorporating contextual information by retrieving and ranking entity notes.
The resulting records retain the institutional descriptions separately as references for subsequent use.

This work makes three main contributions: a curated multimodal dataset of 60{,}397 Europeana records with associated images, catalog fields, and contextual notes, while preserving the original records and their provenance; a documented construction pipeline with training, validation, and test splits designed to prevent overlap among related records; and a characterization of the dataset's composition, languages, and catalog-field coverage across collections.

The following sections review related work (Section~\ref{sec:related}), describe CHICO and its construction (Section~\ref{sec:chico}), present its availability (Section~\ref{sec:release}), discuss limitations (Section~\ref{sec:limitations}), and provide concluding remarks (Section~\ref{sec:conclusion}).

\section{Related Work}
\label{sec:related}

\paragraph{Datasets and Multimodal Approaches in Cultural Heritage.}
Existing resources support vision--language research on cultural heritage items, addressing different tasks and annotation formats.
DEArt provides European art images with object and pose annotations for detection and scene understanding~\citep{reshetnikov-etal-2023-deart}.
EMA introduces a Brazilian cultural heritage image dataset with annotated images from museum collections, supporting automated metadata annotation through image classification~\cite{de2022ema}.
For artwork captioning, \citet{cioni-etal-2023-diffusion} explore diffusion-based data augmentation, while \citet{castellano-etal-2024-artwork} combine vision--language models and language-based methods to generate and refine textual descriptions.
\citet{bongini-etal-2023-gpt3vqa} investigate visual question answering by combining visual information with contextual knowledge about artworks.
Other approaches explore multimodal retrieval of iconographic concepts using the Iconclass classification system~\citep{banar-etal-2021-iconclass,santini-etal-2023-iconclass}.

\paragraph{Metadata Aggregation and Digital Cataloging.}
Digital aggregation initiatives bring together cultural heritage records from different institutions, reflecting diverse cataloging practices and metadata structures.
Europeana and Brasiliana Museus illustrate different approaches to aggregating and providing access to digital cultural heritage collections~\cite{isaac2013europeana,siqueira2022brasiliana}.
Europeana uses the Europeana Data Model (EDM) to represent aggregated records, incorporating Dublin Core metadata terms and linked entities~\cite{isaac2013europeana,dcmi-terms}.
In Brazil, the Tainacan platform supports the management and publication of digital museum collections, while Brasiliana Museus uses metadata mapping and aggregation to bring together collections from different institutions~\cite{siqueira-etal-2022-tainacan}.

Research involving these resources has explored approaches to semantic enrichment and computational cataloging.
\citet{koch-etal-2021-chia} investigate the use of semantic vocabularies and image analysis to improve the description and retrieval of historical images, while \citet{domingo-etal-2020-shabtis} combine visual recognition with semantic integration into Europeana.
\citet{abele-etal-2026-archivegpt} examine vision--language models for archival image cataloging, comparing generated descriptions with human-authored texts.
Institutional descriptions provide valuable curatorial knowledge, although their content may extend beyond the evidence available in associated metadata or images~\citep{maleve-2021-ruins,hanley-etal-2021-alttext}.

CHICO takes cultural heritage records as the unit of analysis, bringing together curated catalog fields, associated images, and contextual evidence while preserving original Europeana provenance.

\section{The CHICO Dataset}
\label{sec:chico}

The integration of catalog metadata, visual information, and contextual sources underpins CHICO's multimodal structure.
Derived from Europeana, the dataset retains the original records alongside curated information, representing these complementary sources.
Its construction involves selecting and processing records, validating associated images, and organizing information for subsequent experimental use.

\subsection{Dataset Construction}
\label{sec:construction}
\label{sec:split}

The dataset was constructed from a local harvest of approximately 33.5 million Europeana records, collected without topical restrictions and filtered for openly licensed content and declared media availability.
Each record was retrieved in its complete Europeana Data Model (EDM) representation and normalized to facilitate data management and subsequent processing.
From this collection, 100{,}000 records were randomly sampled, without stratification by provider, language, or item type.

The final research sample was also selected by verifying the usability of associated images.
Media files were downloaded and checked for successful decoding as raster images.
Only records with at least one successfully decoded image were retained, yielding 60{,}397 records with 83{,}852 images from 74 providers (84.8\% single-image; 15.2\% multi-image).
This validation step also influenced the sample composition, as collections whose media links resolve to decodable images were more likely to be retained than those linking to viewer pages or packaged scans.

\begin{figure*}[t]
    \centering
    \includegraphics[width=\textwidth]{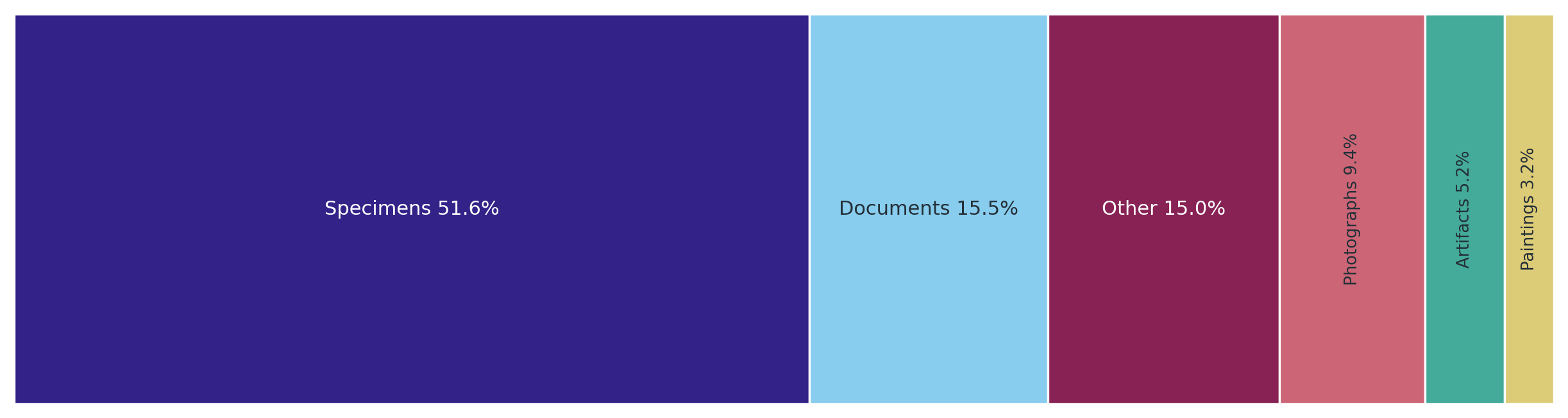}
    \caption{Distribution of object categories across the 60{,}397 records.}
    \label{fig:composition}
\end{figure*}

To characterize the dataset's composition, we classified records into broad item categories based on their cataloged types, considering both source-language and English representations (Figure~\ref{fig:composition}).
Natural-history specimens form the largest group and already represented approximately 40\% of the initial random sample; their share rises after image validation because these providers more frequently expose media links that resolve to decodable images.
Consequently, the resulting set is a fixed, image-validated research sample rather than a balanced or representative sample of the broader Europeana collection.

The catalog records are multilingual, reflecting the diversity of the contributing collections.
English is the most frequently declared catalog language, alongside substantial non-English material.
Catalog fields are preserved in their source languages, with English representations included when available.

To minimize information leakage across experimental subsets, we grouped related records into connected components and assigned them together to training (80\%), validation (10\%), and test (10\%) partitions.

\subsection{Multimodal Record Structure}
\label{sec:record-structure}

CHICO's multimodal structure brings together catalog metadata, associated images, and contextual information for each cultural heritage item.
This structure is defined as
\begin{equation}
\label{eq:evidence}
E = \bigl(M,\, I,\, R,\, d^{\mathrm{ref}}\bigr),
\end{equation}
where \(M\) contains curated catalog fields, \(I\) represents the associated images and their roles, \(R\) contains contextual notes selected from related entities referenced by the record, and \(d^{\mathrm{ref}}\) denotes the institutional description when available.
The original Europeana record is also preserved, maintaining access to the source metadata.
For evidence-grounded description generation, \(M\), \(I\), and \(R\) constitute the designated supporting evidence, while \(d^{\mathrm{ref}}\) is retained separately for comparison and excluded from generation inputs.

\begin{figure*}[t]
    \centering
    \includegraphics[width=\textwidth]{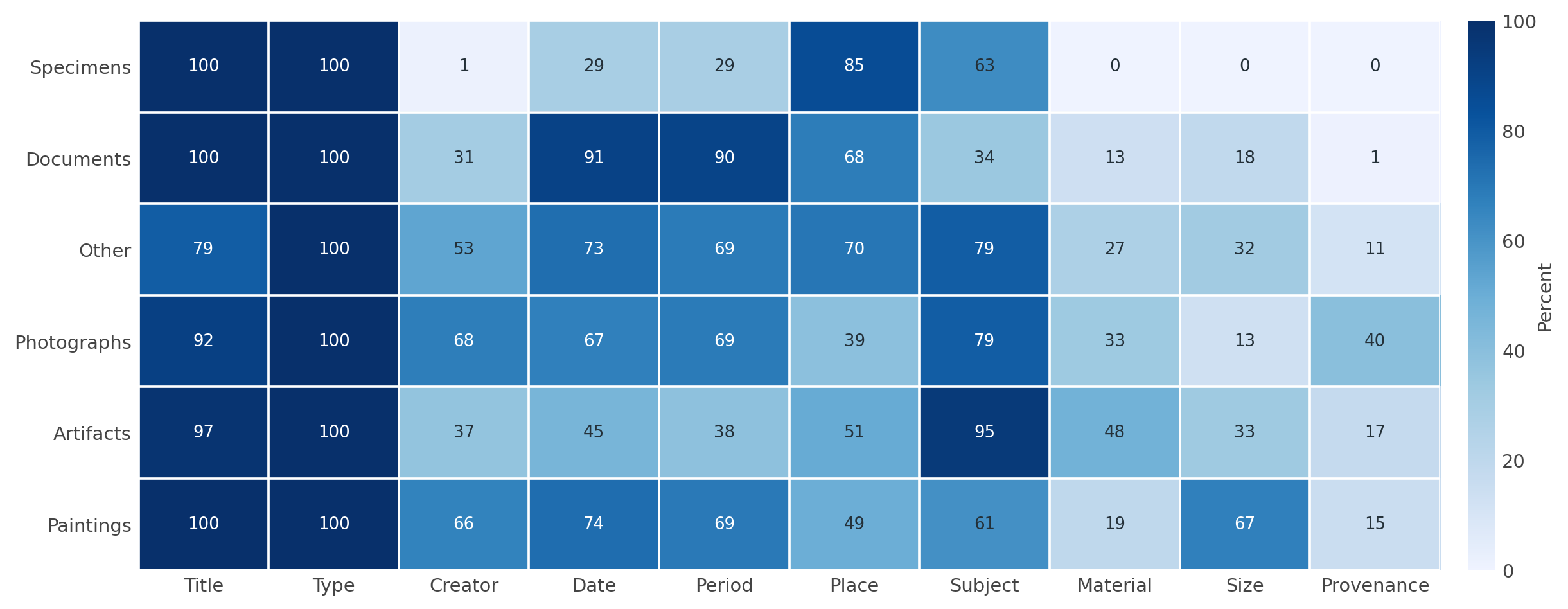}\\[0.2em]
    {\small (a)}\\[0.9em]
    \includegraphics[width=0.78\textwidth]{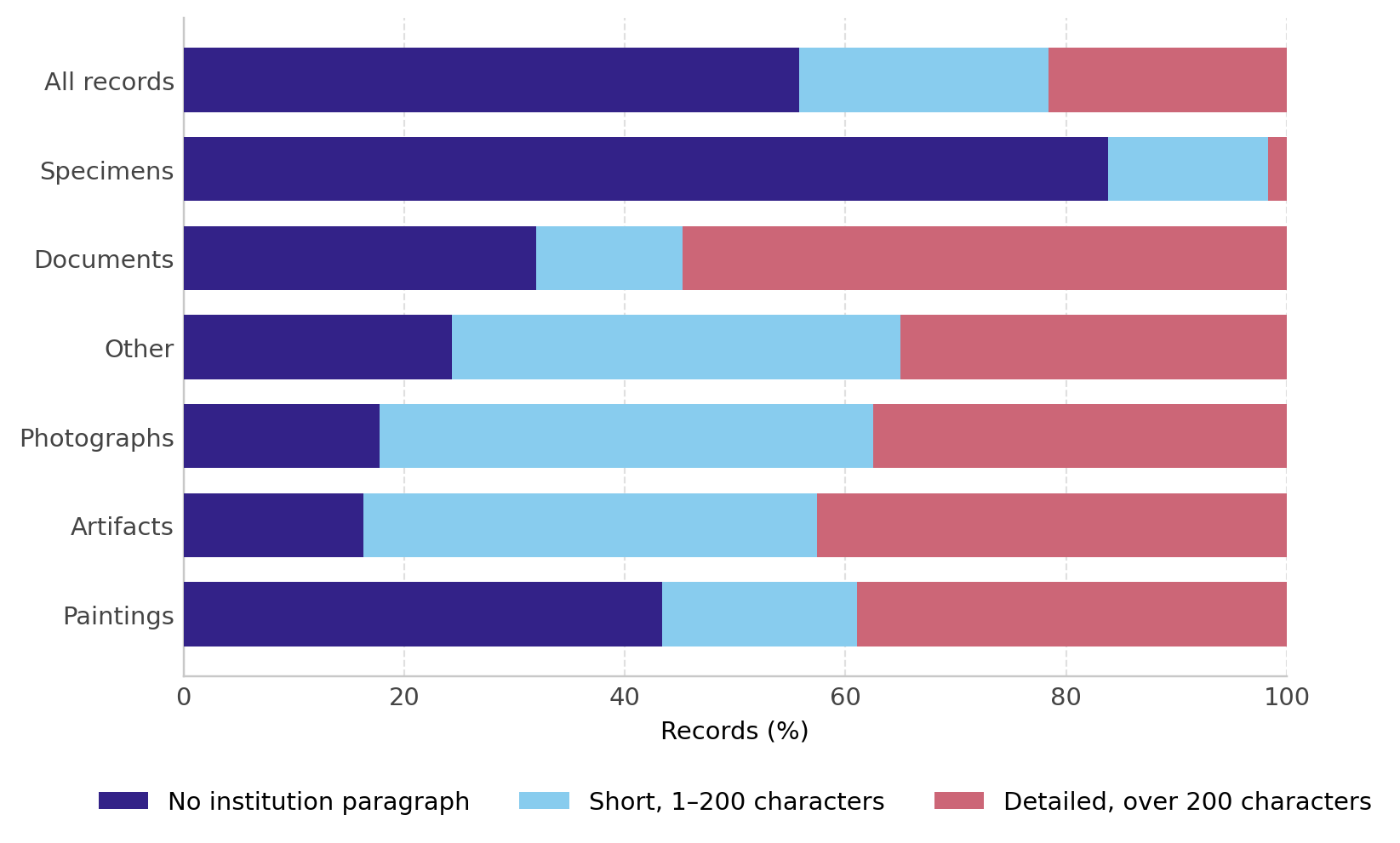}\\[0.2em]
    {\small (b)}
    \caption{(a)~Coverage of curated catalog fields by object category. (b)~Institutional descriptions (\(d^{\mathrm{ref}}\)) by object category.}
    \label{fig:characterize}
\end{figure*}

The selection of catalog fields was informed by an examination of 1{,}000 randomly sampled Europeana records, considering field availability, multilingual representations, and their relevance to describing cultural heritage items~\citep{isaac2013europeana}.
This analysis guided the definition of the curated Cultural Heritage Object (CHO) fields, prioritizing information associated with the items themselves rather than administrative metadata.
Field selection considered descriptive relevance alongside availability, retaining information such as provenance, material, and extent even when less frequently recorded.
The object-level fields in \(M\) include title, type, creator, date, period, place, subject, material, size, and ownership history.
The distribution of catalog information differs considerably across cultural heritage categories (Figure~\ref{fig:characterize}a).

Associated images (\(I\)) are assigned roles that distinguish object views from details, labels, mounts, and ambiguous frames.
Records with a single image are treated as object views; records with multiple images are labeled automatically so that later use can separate appearance evidence from text-bearing or non-object frames.

Contextual information (\(R\)) comes from descriptive notes of related entities referenced by the record---concepts, places, agents, and timespans---rather than from the curated catalog fields in \(M\).
Within each record, candidate notes are retrieved with BGE-M3 and reranked with BGE-reranker-v2-m3~\cite{chen-etal-2024-m3-embedding}, guided by \(M\), so that more relevant notes are ordered first.
The ranked list is then filtered to avoid repeating information already stated in those fields; when no suitable notes remain, \(R\) is left empty.

Institutional descriptions (\(d^{\mathrm{ref}}\)) vary across categories in both presence and length (Figure~\ref{fig:characterize}b).
They are therefore not treated as ground-truth object descriptions: the field is frequently empty, and when present, it does not reliably reflect a complete account of the object.

\section{Availability}
\label{sec:release}

CHICO is publicly available for research use.
Construction and analysis code is released in the heritage-data-hub repository.\footnote{\url{https://github.com/AI-Unicamp/heritage-data-hub}}
The dataset is deposited on Zenodo~\cite{oliveira2026chico}.
The deposit includes the catalog records, train/validation/test splits, image URLs with checksums for verification, and documentation.
An optional archive provides the contextual enrichment notes.
Image files are not redistributed; they can be retrieved from the recorded URLs and checked against the provided checksums, subject to each record's copyright and reuse terms.

\section{Limitations}
\label{sec:limitations}

As a fixed research sample derived from openly licensed Europeana records with associated media, the dataset should not be considered representative of the broader Europeana collection.
The requirement for successfully decoded images influences the sample composition, favoring providers whose media links resolve to decodable images over those that lead to viewer pages, packaged scans, or unavailable content.

The dataset partitions link related records through exact matches of identifiers, image URLs, and cryptographic hashes.
Visually similar images already present in the collection typically receive different hashes and are therefore not treated as related, so such near-duplicates may still appear across splits.
Additionally, the broad item categories used in the dataset analysis are derived from heterogeneous catalog type values and may not fully represent the original classifications.
Contextual enrichment draws only on notes of related entities referenced by each record.

Although CHICO was originally developed to support evidence-grounded description generation, this release focuses on dataset construction and characterization.
Its potential use in other applications, including multimodal retrieval and metadata enrichment, remains to be evaluated.

\section{Conclusion}
\label{sec:conclusion}

Sparse and heterogeneous catalog records in cultural heritage aggregators make it difficult to assemble multimodal evidence for reproducible research.
This work addresses that gap with a fixed research sample of 60{,}397 Europeana records that brings together curated catalog fields, verified images, and contextual notes while preserving the original records and their provenance.
The construction yields predefined splits that group related records and a multimodal structure that retains the institutional description separately for comparison.
The dataset is characterized by its composition, languages, and catalog-field coverage across collections, along with documentation of its construction.
Code and data are publicly available, providing a reusable foundation for evidence-grounded description generation and related multimodal research on cultural heritage collections.

\section*{Acknowledgments}
This work was partially funded by the Coordenação de Aperfeiçoamento de Pessoal de Nível Superior -- Brasil (CAPES) -- Finance Code 001.
The authors are also affiliated with the Artificial Intelligence Lab, Recod.ai.

\bibliography{custom}

\end{document}